\documentclass[floatfix,twocolumn,amsmath,superscriptaddress,showkeys]{revtex4}

\pdfoutput=1

\usepackage{color}
\usepackage{xspace}
\usepackage{amssymb}

\renewcommand{\in}{{\rm in}}
\newcommand{\Acell}{\ensuremath{A_{\rm cell}}\xspace}
\newcommand{\Aoff}{\ensuremath{A_{\rm off}}\xspace}
\newcommand{\eTmin}{\ensuremath{{e_{T\rm{ min}}}}\xspace}
\newcommand{\microm}{\ensuremath{\mu\text{m}}\xspace}

\newcommand{\ourtitle}{A biophysical model of cell adhesion mediated by immunoadhesin drugs and antibodies}

\usepackage{graphicx}
\usepackage{subfigure}

\begin{document}

\title{A biophysical model of cell adhesion mediated by\\ immunoadhesin drugs and antibodies}

\author{Ryan N. Gutenkunst}
\email{ryan@gutenkunst.org}
\affiliation{Theoretical Biology and Biophysics Group, Theoretical Division, Los Alamos National Laboratory, Los Alamos, NM 87545}
\affiliation{Center for Nonlinear Studies, Theoretical Division, Los Alamos National Laboratory, Los Alamos, NM 87545}
\author{Daniel Coombs}
\affiliation{Department of Mathematics and Institute of Applied Mathematics, University of British Columbia,  Vancouver, British Columbia V6T 1Z2, Canada}
\author{Toby Starr}
\thanks{Deceased February 29, 2008}
\affiliation{Department of Pathology, New York University School of Medicine and Program in Molecular Pathogenesis, Skirball Institute of Biomolecular Medicine, 540 First Avenue, New York, NY 10016}
\author{Michael L. Dustin}
\affiliation{Department of Pathology, New York University School of Medicine and Program in Molecular Pathogenesis, Skirball Institute of Biomolecular Medicine, 540 First Avenue, New York, NY 10016}
\author{Byron Goldstein}
\affiliation{Theoretical Biology and Biophysics Group, Theoretical Division, Los Alamos National Laboratory, Los Alamos, NM 87545}

\begin{abstract}

A promising direction in drug development is to exploit the ability of natural killer cells to kill antibody-labeled target cells.
Monoclonal antibodies and drugs designed to elicit this effect typically bind cell-surface epitopes that are overexpressed on target cells but also present on other cells.
Thus it is important to understand adhesion of cells by antibodies and similar molecules.
We present an equilibrium model of such adhesion, incorporating heterogeneity in target cell epitope density and epitope immobility.
We compare with experiments on the adhesion of Jurkat T cells to bilayers containing the relevant natural killer cell receptor, with adhesion mediated by the drug alefacept.
We show that a model in which all target cell epitopes are mobile and available is inconsistent with the data, suggesting that more complex mechanisms are at work.
We hypothesize that the immobile epitope fraction may change with cell adhesion, and we find that such a model is more consistent with the data.
We also quantitatively describe the parameter space in which binding occurs.
Our results point toward mechanisms relating epitope immobility to cell adhesion and offer insight into the activity of an important class of drugs.

\end{abstract}

\keywords{Alefacept, CD2, CD58, Fc$\gamma$RIII, Immobility}

\maketitle

\section{Introduction}

When a pathogen elicits a humoral immune response, antibodies are produced that bind to specific epitopes on the surface of the pathogen.
Once antibodies have bound to  the pathogen it is labeled as foreign, and various processes can follow that lead to its elimination.
One such process, antibody-dependent cell-mediated cytotoxicity (ADCC), involves natural killer (NK) cells binding through their Fc$\gamma$RIIIa (CD16a) receptors to IgG antibodies decorating the pathogen (reviewed in~\cite{bib:Ianello2005}).
The coupling of an NK cell to a target cell brings parts of the surfaces of the two cells into proximity, within roughly 100 \AA.
In the region of tight contact where antibodies form bridges between the two cells, both the density of epitopes on the target cell and the density of Fc receptors on the NK cell are locally increased.  
When the density of Fc receptors in the contact region on the NK cell is sufficiently high, a cellular response is triggered, the end point of which  is the release of lytic granules containing perforin and granzymes, whose combined effect results in the killing of the target cell~\cite{bib:daSilva2002, bib:Trapani2002, bib:Bryceson2005}.
Depending on the nature of the epitope and type of cell,  the aggregation of epitopes  on the target cell may also trigger cellular responses~\cite{bib:Ghetie1997,bib:Friedman2005}.

Monoclonal antibodies and antibody-like fusion proteins have been developed to take advantage of ADCC.
These drugs target naturally occurring proteins that are overexpressed on  tumor cells and on populations of cells that drive autoimmune responses~\cite{bib:Clynes2000,bib:Waldmann2003,bib:Ianello2005,bib:Arnould2006,bib:Koon2006}. 
Unfortunately the drug will also target a subset of healthy cells because the target is a naturally occurring protein.
An obvious question, which we address in this paper, is what properties of the drug, the cells that express the target protein, and the NK cells determine the drug's ability to discriminate between pathogenic and healthy cells?
A second question that we consider, that is closely related to the first, is what  determines the range of drug concentrations over which the drug will couple target cells to NK cells?
These drugs, either in animal models or patients, must compete for Fc receptors on NK cells with endogenous IgG~\cite{bib:Preithner2006}.
We therefore also look at how background IgG influences the range of drug concentrations over which adhesion occurs.

We previously presented an equilibrium model that describes the coupling via a monoclonal antibody (or an appropriate fusion protein) of identical target cells to a surface expressing mobile Fc receptors~\cite{bib:Dustin2007}. 
Here, we significantly extend our model to allow for a target cell population with a heterogeneous surface epitope density.
This allows us to analyze experiments where the percentage of target cells bound is determined as a function of the ligand concentration.
We also extend the model to address the possibility that some fraction of the target epitopes are immobile, including cases in which the immobile fraction depends on epitope cross-linking or the size of the contact region.
These cases model some potential target cell responses to adhesion.

To test  predictions of the model we use an experimental system consisting of a planar  bilayer containing mobile Fc$\gamma$RIIIb (CD16b) receptors,  Jurkat T cells expressing the cell-adhesion molecule CD2, and a drug, alefacept, that binds the target cell to the bilayer~\cite{bib:Dustin2007}.
Fc$\gamma$RIIIb differs from Fc$\gamma$RIIIa, the receptor on NK cells, in that it  lacks a transmembrane region and a cytoplasmic tail and anchors to membranes via glycosolphosphatidylinositol~\cite{bib:Scallon1989}.
Further, the extracellular domains of the two receptors differ by six amino acids, which probably accounts for  Fc$\gamma$RIIIb having a lower affinity for IgG than Fc$\gamma$RIIIa~\cite{bib:Scallon1989,bib:Anderson1990}.
Alefacept is a recombinant fusion protein that  has an antibody-like architecture where the Fab binding sites have been replaced by the natural ligand for CD2, the extracellular domain of CD58~\cite{bib:Dustin1991,bib:Dustin1997}, fused to the human IgG1 hinge, C$_H$2, and C$_H$3 domains~\cite{bib:daSilva2002}.
It is used in the treatment of psoriasis, an autoimmune disease, where it targets memory-effector T cells that have increased levels of CD2 on their surface.
Alefacept reduces the number of circulating memory-effector T cells in treated patients and  mediates ADCC in vitro~\cite{bib:Majeau1994, bib:Ellis2001, bib:daSilva2002, bib:Cooper2003, bib:Chamian2005}.

Alefacept is an example of an immunoadhesin, which is molecule that uses the basic framework of an IgG antibody, but replaces the Fab binding sites with the ecotodomain of an adhesion molecule.
Immunoadhesins have the specificity of an adhesion molecule as well as some properties of an antibody, such as the ability to bind to Fc receptors and a half-life in plasma that is similar to IgG~\cite{bib:Byrn1990,bib:Ashkenazi1995}.
An interesting property of  alefacept is that it mediates adhesion and killing of target cells by NK cells at nM concentrations~\cite{bib:Dustin2007} even though both the binding of IgG to Fc$\gamma$RIIIa and the binding of CD58 to CD2~\cite{bib:Dustin1996} are low affinity, in the $\mu$M range.
The model we present will show how the range of drug concentrations  over which adhesion occurs depends on these equilibrium constants as well as the other parameters of the system.

\section{Model and methods}

We consider a population of target cells expressing a particular epitope, with some fraction of the epitopes freely diffusing in the target cell membrane and the remainder immobile, i.e.\ fixed in position on the membrane.
Additionally, we consider a bilayer with mobile receptors diffusing on its surface and a ligand capable of simultaneously binding both the epitope and the receptor through different sites.
The ligand is either a monoclonal antibody or an immunoadhesin; its Fab arms bind monovalently or divalently to the epitope on the target cells, and its Fc leg binds monovalently to the receptor on the bilayer.
At some ligand concentration a contact region forms between the cell and the bilayer; its area is an increasing function of the number of ligand-mediated bridging bonds that form.

\subsection{Concentrations and equilibrium constants}

\begin{figure}
\centering
\includegraphics[]{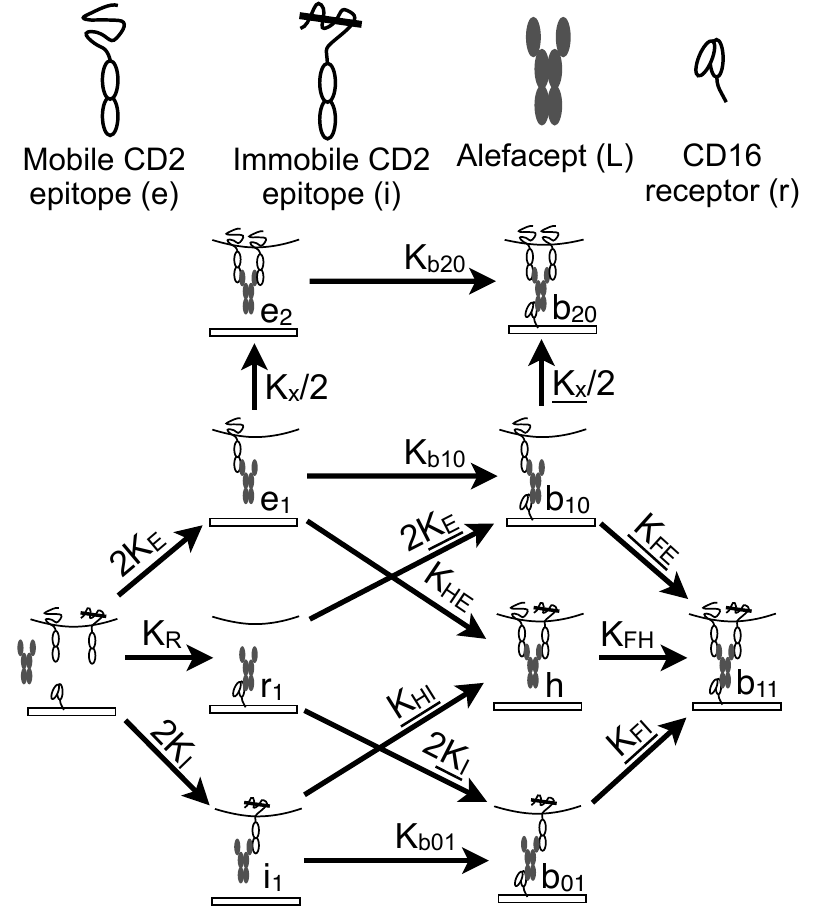}\\
\caption{Model reaction network. All molecular species and reactions are labeled. 
All reactions are reversible; the arrow in the figure denotes the forward direction for defining the equilibrium constant, which labels the arrow. Underlined rate constants are eliminated using detailed balance.\label{fig:rxns}}
\end{figure}

The potential reactions between epitopes, ligand, and receptor are illustrated in Fig.~\ref{fig:rxns}.
Each molecular complex is labeled by our mathematical notation for its surface concentration.
All species except those involving a bridging bond ($b_{10}$, $b_{20}$, $b_{11}$, and $b_{01}$) exist both inside and outside the contact region, and the subscript `in' denotes species inside the contact region.
Detailed balance places six constraints on the equilibrium constants, which we use to eliminate the underlined constants in Fig.~\ref{fig:rxns} (see Supporting Material for more detail.)
To find the equilibrium state of this system for any given bulk ligand concentration $L$, we solve five algebraic equations for five unknowns: the free immobile epitope concentration outside $i$ and inside $i_\in$ the contact region, the free mobile epitope $e$ and receptor $r$ concentrations outside the contact region, and the fraction of the target cell surface $\delta$ comprising the contact region.
To make analysis tractable, we make several simplifying assumptions regarding the equilibrium configuration of receptors and epitopes.

Our first assumption is that the equilibrium constants for reactions involving immobile epitopes are identical to the corresponding constants involving mobile epitopes: $K_I = K_E$, $K_{HE} = K_x$, $K_{b01} = K_{b10} \equiv K_{b1}$, and $K_{FH} = K_{b20} \equiv K_{b2}$.
Making this assumption substantially reduces the number of unknown parameters.
We expect that this assumption leads to negligible error, because the relevant physical interactions are identical for mobile and immobile epitopes.

Our second assumption is that the typical distance between immobile epitopes on the target cell is large compared with the span of the two arms of the ligand, so that the ligand cannot cross-link immobile epitopes.
Thus we do not consider complexes containing more than one immobile receptor.
Given a CD2 surface density $\rho$, and assuming the CD2 epitopes are uniformly distributed, the probability $P(d < a)$ that an epitope's nearest neighbor is a distance $a$ away or closer is~\cite{bib:Chandrasekhar1943}:
\begin{equation}\label{eqn:distances}
P(d < a) = 1 - e^{-\pi \rho a^2}.
\end{equation}
As detailed later, each T-cell contains of order 64,000 CD2 epitopes, over a surface area of roughly 800 $\microm^2$, yielding a density of $\rho = 80\,\microm^{-2}$.
Given this density, the probability that an epitope's nearest neighbor is closer than the span of roughly 10 nm~\cite{bib:Harris1992} between epitope binding arms of an antibody-like molecule, such as alefacept, is less than 3\%.
We expect that cross-links between immobile epitopes will indeed be rare because the density of immobile epitopes is even lower than the total epitope surface density.

Our third assumption relates the free mobile epitope and receptor concentrations inside and outside the contact region.
In earlier experiments, fluorescently labeled CD48 (Cy5-CD48) was coupled to the bilayer, and it was observed that the fluorescence from CD48 was reduced in the contact region to approximately 75\% of its value outside the contact region~\cite{bib:Dustin2007}.
This suggests that the contact region introduces steric hindrance and partitions mobile surface proteins between the inside and outside of the contact region.
We assume that at equilibrium $e_\in = \sigma_E e$ and $r_\in = \sigma_R r$,
where $e_\in$ and $r_\in$ are the free epitope and free receptor concentrations inside the contact region and  $\sigma_E$ and $\sigma_R$ are equilibrium partition coefficients.
The partition coefficients for Fc$\gamma$RIIIb and CD2 have not been determined; we assume they behave similarly to CD48 because they are of similar size, so we take $\sigma_R = \sigma_E = 0.75$.  

Using the law of mass action and these additional assumptions, we can write down the equilibrium concentration of all bound complexes in terms of the free epitope, receptor, and ligand concentrations.
For example, the concentration $h_\in$ of complexes inside the contact region consisting of a ligand cross-linking a mobile and an immobile epitope is:
\begin{equation}\label{eqn:hin}
h_\in = K_x i_\in e_{1\in} = 2 K_x K_E L i_\in e_\in = 2 K_x K_E L \sigma_E i_\in e .
\end{equation}
The factor of 2 in calculating the concentration $e_{1\in}$ of complexes between a ligand and mobile epitope arises because $K_E$ is a single-site equilibrium constant, and there are two potential binding sites on the ligand.
Similarly, the concentration $b_{11}$ of bridging complexes involving a receptor, a mobile epitope, and an immobile epitope is:
\begin{equation}\label{eqn:b11}
b_{11} = K_{b2} h_\in r_\in = K_{b2} h_\in \sigma_R r.
\end{equation}
The full system of equilibrium relations is given in Supporting Material.

Given our assumptions, in the limit that target cells are sparse the equilibrium state will depend on five equilibrium constants ($K_R$, $K_E$, $K_x$, $K_{b1}$, and $K_{b2}$), two partition coefficients ($\sigma_E$ and $\sigma_R$), the total receptor $r_T$, epitope $e_T$, and ligand concentrations $L_T$, the epitope immobile fraction $\eta$, and a parameter $\beta$ that relates the area of the contact region to the number of bridging bonds.
In addition, to connect our model with the data, we require the surface area \Acell of the Jurkat T cells studied, and an offset area \Aoff that accounts for non-specific adhesion between the cells and the bilayer (detailed later).
In our analyses, a number of these parameters were held fixed (Table~\ref{tbl:fixed_params}).

Most of our fixed parameter values come directly from measurements; an exception is $K_x$, the cross-linking constant for alefacept binding.
To estimate $K_x$, we equate the measured  apparent dissociation constant $K_D$ for alefacept adhering to T cells to the inverse of the initial slope $m_0$ of a Scatchard plot for a bivalent ligand binding to a monovalent receptor~\cite{bib:Wofsy1992}:
\begin{equation}
m_0 = -\frac{2 K_E (1 + K_x e_T)^2}{1+K_x e_T/2} = -\frac{1}{K_D}.
\end{equation}
Using the mean CD2 count measured for our cells of $6.4 \times 10^4$ and $A_{cell}$ to calculate $e_T$, along with $K_D$ = 100 nM~\cite{bib:Dustin2007} yields the value for $K_x$ in Table~\ref{tbl:fixed_params}.
With the exception of the fit parameter $K_{b2}$, our results are insensitive to the precise value of $K_x$ (Fig.~\ref{fig:changing_Kx} in Supporting Material).
The parameter \Aoff is inferred from our data, as discussed below.

\begin{table}
\caption{Fixed parameter values\label{tbl:fixed_params}}
\begin{tabular*}{\columnwidth}{@{\extracolsep{\fill}} ccc}
parameter & value & reference\\
\hline
$K_R$ & $1.0\times 10^6$ M$^{-1}$ & \cite{bib:Galon1997,bib:Chesla2000,bib:Maenaka2001}\\ 
$K_E$ & $6.7\times 10^5$ M$^{-1}$ & \cite{bib:Dustin1997}\\
$K_x$ & $4.5\times 10^{-2}$ $\microm^{2}$ & this work\\
$\sigma_E$ & 0.75 & \cite{bib:Dustin2007}\\
$\sigma_R$ & 0.75 & \cite{bib:Dustin2007}\\
\Acell & 800 $\microm^2$ & \cite{bib:Dustin2007}\\
\Aoff & 6.2 $\microm^2$ & this work\\
\end{tabular*}
\end{table}

\subsection{Conservation laws}

In the experiments we consider, there is negligible depletion of ligand so the free ligand concentration is well approximated by the total ligand concentration ($L \approx L_T$).
Conservation of epitopes and receptors, however, introduces additional constraints on the concentration of various complexes.

In our model, there are three classes of epitopes: mobile epitopes, immobile epitopes outside the contact region, and immobile epitopes inside the contact region. We assume that the concentrations of all species have reached equilibrium.
For mobile epitopes, we have
\begin{multline}
\Acell (1-\eta) e_T = (\Acell - A)( e + e_1 + 2 e_2 + h)\\+A(b_{10} + 2 b_{20} + b_{11} + e_\in +e_{1{\rm in}}+2e_{2{\rm in}} + h_\in), 
\end{multline}
where $e_T$ is the average epitope density on the cell surface, equal to the total epitope count $E_T$ divided by the cell area $\Acell$, and $A$ is the area of the contact region.
This equation expresses the fact that the total number of mobile epitopes (left-hand side) must be equal to the total number in complexes outside and inside the contact region.
In terms of the fraction $\delta$ of the cell surface in the contact region, the above conservation law is:
\begin{multline}\label{eqn:epitope_constraint}
(1-\eta) e_T = (1 - \delta)( e + e_1 + 2 e_2 + h) \\+\delta (b_{10} + 2 b_{20} + b_{11} + e_\in +e_{1{\rm in}}+2e_{2{\rm in}} + h_\in).
\end{multline}
Similarly, for immobile epitopes outside the contact region we have
\begin{equation}\label{eqn:i_constraint}
\eta \, e_T = i + i_1 + h,
\end{equation}
and for immobile epitopes inside the contact region we have
\begin{equation}\label{eqn:iin_constraint}
\eta \, e_T = i_\in + i_{1\in} + h_\in + b_{01} + b_{11}.
\end{equation}

For receptors in the bilayer, we have
\begin{multline}
A_{bl} \, r_T = A (r_\in + r_{1\in} + b_{10} + b_{20} + b_{01} + b_{11})\\ + (A_{bl} - A)(r + r_1),
\end{multline}
where $A_{bl}$ is the total area of the bilayer divided by the number of adhered cells.
Dividing by \Acell and rearranging yields
\begin{multline}\label{eqn:receptor_constraint}
r_T = \alpha \delta (r_\in + r_{1\in} + b_{10} + b_{20} + b_{01} + b_{11})\\ + (1 - \alpha \delta)(r + r_1),
\end{multline}
where $\alpha = \Acell/A_{bl}$.
In the experiments we analyze, for all ligand concentrations the adhered cells are sparsely distributed over the bilayer, so we take $\alpha = 0$.

\subsection{Contact region growth law}
Bell, Dembo, and Bongrand argued~\cite{bib:Bell1984} that the bridging bond density between two adhered cells is determined by a constant repulsive pressure arising from electrostatic repulsion caused by negative charges associated with cell surfaces and steric stabilization effects.
The steric effects arise because cell membranes are coated by a hydrated layer of long-chain polymers (glycocalyces) that must compress as cells are brought together and water is squeezed out of the contact region.
Together with the assumption that cells are easily deformed, this argument implies that the area of the contact region grows linearly with the number of bridging bonds.

Although we expect the repulsion between our target cells and bilayer to be smaller than that between two cells, we expect the repulsive forces to be of similar origin.
Moreover, in our experiments the small contact regions observed cause only small cellular deformations, and our target Jurkat T cells are significantly more easily deformed than some other cell types, such as neutrophils and HL60 cells~\cite{bib:Rosenbluth2006}.
However, in our equilibrium data, the observed average area of the contact region $\left<A_{\rm obs}\right>$ does not go to zero as the number of bridging bonds goes to zero (Fig.~\ref{fig:hetero_fits}A), an effect seen previously in the binding of Jurkat T cells to bilayers containing the natural CD2 ligand CD58~\cite{bib:Shao2005}.
We hypothesize that some additional non-specific adhesion occurs after cells settle onto the membrane.
We account for this effect by subtracting an offset area \Aoff from our data, so that in our modeling we compare with the average specific adhered area $\left<A\right>$, which is $\left<A_{\rm obs}\right>  - \Aoff$.
Our procedure for estimating \Aoff is discussed in the Results section.
Once we have subtracted the offset, we observe a linear relationship between average contact area and average bridging bond number, suggesting that the bond density $\beta$ is indeed a constant.
In addition to the constraints from conservation of epitopes and receptors, we thus have an additional constraint on bond density:
\begin{equation}\label{eqn:beta_constraint}
\beta = b_{10} + b_{20} + b_{01} + b_{11}.
\end{equation}

Solving the five constraint equations (Eq.~\ref{eqn:epitope_constraint}, \ref{eqn:i_constraint}, \ref{eqn:iin_constraint},  \ref{eqn:receptor_constraint}, and \ref{eqn:beta_constraint}) for the five unknowns $e$, $i$, $i_\in$, $r$, and $\delta$ allows us to calculate the area of the contact region for a cell with a specified epitope density $e_T$ given the ligand concentration $L$.
We solve the constraint equations and perform all numerics using the Python library SciPy~\cite{bib:SciPy}.
Uncertainties on fit parameters are calculated via bootstrapping, with over 250 bootstrap data sets for each model.

\subsection{Heterogeneous  density of epitopes on target cells}

We now consider a target cell population with a normalized distribution $f(e_T)$ of epitope densities.
For a given ligand concentration, if \eTmin is the minimum epitope density at which adhesion will occur, the fraction of cells bound to the bilayer is
\begin{equation}
\text{fraction bound} = \int_\eTmin^\infty f(e_T)\,de_T,
\end{equation}
and the average area of the specific contact region is
\begin{equation}
\left<A\right> = \Acell \left<\delta\right> = \frac{\int_\eTmin^\infty \delta(e_T) f(e_T)\,de_T}{\int_\eTmin^\infty f(e_T)\,de_T}.
\end{equation}

We will primarily model the distribution of target cell epitope densities by the three parameter Weibull distribution, which has density
\begin{equation}\label{Weibull}
f(x; \gamma, k, \lambda)=
\begin{cases} \frac{k}{\lambda} \Big( \frac {x-\gamma}{\lambda}\Big)^{k-1}e^{-( \frac {x-\gamma}{\lambda})^k} & \text{if } x>\gamma\\ 0 & \text{if } x \le\gamma.
\end{cases}
\end{equation}
In our fits, we fix the location parameter $\gamma$ to zero.
The parameter $k$ is the Weibull shape parameter.
For $k<2.6$, $f(x)$ is skewed to the right, for $2.6<k<3.7$ it is essentially unskewed and looks like a normal distribution, and for $k>3.7$ it is skewed to the left.
For $k=1$ and $\gamma=0$ the Weibull distribution reduces to the exponential distribution.

\begin{figure}
\centering
\includegraphics[]{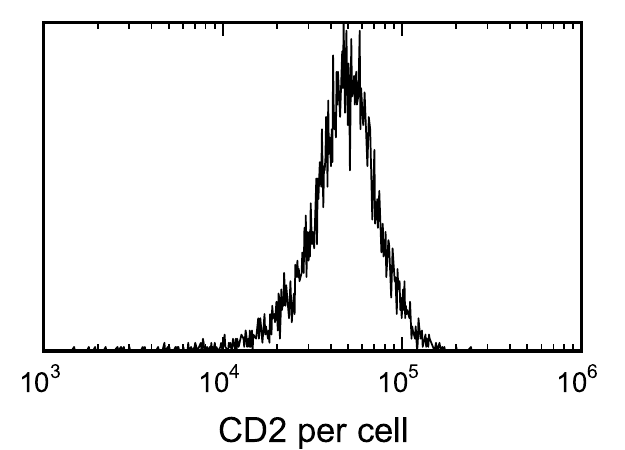}
\caption{Distribution of CD2 epitope count on Jurkat T cells like those used in our experiments, as determined by flow cytometry.\label{fig:flowcyt}}
\end{figure}

Shown in Fig.~\ref{fig:flowcyt} is the distribution of expression of CD2 on the surfaces of a population of Jurkat T cells, the target cells in our study, as determined by flow cytometry.
This distribution serves as a baseline to judge the distributions arising from our model fits.
These measurements were performed on a Becton Dickinson FacsCaliber.
Fluorescein isothiocyanate (FITC) labeling of antibodies and the determination of the fluoresceine:protein ratio was determined using absorption spectroscopy~\cite{bib:Wells1966}.
Calibration was performed using FITC standard beads obtained from Bangs Laboratories (Fishers, IN).
Antibody-stained cells were washed twice prior to analysis.

\subsection{Analysis of the model with fully mobile epitopes}\label{sec:mobile}

When all epitopes are mobile ($\eta = 0$) and there is no depletion of bilayer receptors ($\alpha = 0$) the model simplifies substantially to a system of three equations for $e$, $r$, and $\delta$.
This reduced system offers several useful analytic results.

\subsubsection{Requirements for adhesion}\label{sec:mobile_adhesion}

For a fixed epitope density, adhesion occurs over a range of soluble ligand concentrations: $L_- \le L \le L_+$.
Similarly, for a fixed ligand concentration, adhesion occurs only above a minimal epitope density \eTmin.
As $L$ or $e_T$ approach these bounds, the contact area approaches some minimal value $A_{\rm min}$ below which there is no adhesion.
We make the approximation that $A_{\rm min} = 0$.
Plugging $\delta = 0$ into Eq.~\ref{eqn:epitope_constraint},~\ref{eqn:receptor_constraint}, and~\ref{eqn:beta_constraint} and substituting our expressions for equilibrium species concentrations (such as Eq.~\ref{eqn:hin}) yields a system of three equations for the unknowns $e$, $r$, and $L$.
For a fixed epitope density $e_T$, this system can be reduced to a single cubic equation in $L$, which is given in Supporting Material.
This cubic equation may have either two positive roots ($L_-$ and $L_+$) or no positive roots (no adhesion irrespective of $L$). 
Similarly, for a fixed ligand concentration $L$, the system of three equations can be reduced to a quadratic equation for $e_T$, the larger root of which is \eTmin (see Supporting Material).

We expect that  taking $\delta=0$ rather than setting it equal to some minimal value  introduces negligible error in our estimates of $L_-$, $L_+$, and \eTmin.
In our experiments adhesion was determined in the absence of flow, based on whether there was accumulation of the receptor CD58 in the contact area.
Even in the presence of weak flows, estimates suggest that few bonds, and thus small contact areas, are needed to maintain adhesion~\cite{bib:Bell1981,bib:Hammer1989}.

\subsubsection{Average contact area}\label{sec:mobile_delta}

When the cells adhered to the bilayer negligibly deplete the receptors so that $\alpha = 0$, and all epitopes are mobile so that $\eta = 0$, then
\begin{equation}
\left<\delta\right>=\delta(\left<e_T\right>)
\end{equation}
where $\left<e_T\right>$ is the average epitope density of adhered cells:
\begin{equation}
\left<e_T\right>=\frac {\int_\eTmin^\infty e_T f(e_T) \, de_T} {\int_\eTmin^\infty f(e_T) \, de_T}.
\end{equation}
To prove this, we first note the constraint imposed by our relation for the bond density (Eq.~\ref{eqn:beta_constraint}) when $\alpha = 0$ and $\eta = 0$.
In this case, after substituting our equilibrium relations, $\beta$ is equal to a function of $e$ and $r$ which does not involve $e_T$.
When $\alpha = 0$, $r$ is simply $r_T/(1+K_R L)$ (from Eq.~\ref{eqn:receptor_constraint}), again not involving $e_T$.
Thus, our relation for the bond density provides a constraint on the free epitope density $e$ that is independent of the total epitope density $e_T$.
In other words, all adhered cells will have the same density of free epitopes and thus of all other complexes, irrespective of their total epitope density $e_T$.

When $\alpha = 0$ and $\eta = 0$, our conservation equation for mobile epitopes (Eq.~\ref{eqn:epitope_constraint}) yields the following expression for $\delta$:
\begin{equation}
\delta = \frac{e_T - (e + e_1 + 2 e_2)}{b_{10} + 2 b_{20} + e_{\in} + e_{1\in} + 2 e_{2\in} - (e + e_1 + 2 e_2)}
\end{equation}
Note that all complex concentrations on the right-hand-side ($e$, $e_1$, $e_2$, $b_{10}$, $b_{20}$, $e_\in$, $e_{1\in}$, and $e_{2\in}$) are functions of $e$ and $r$ only, which we have just shown are independent of $e_T$.
Thus, taking the average over adhered cells, we have
\begin{equation}
\left<\delta\right> = \frac{\left<e_T\right> - (e + e_1 + 2 e_2)}{b_{10} + 2 b_{20} + e_{\in} + e_{1\in} + 2 e_{2\in} - (e + e_1 + 2 e_2)},
\end{equation}
where the right-hand-side is simply the specific contact area calculated using the average epitope density of adhered cells, $\left<e_T\right>$.
Thus, for the case in which $\alpha = 0$ and all receptors are mobile, calculating $\left<\delta\right>$ does not require explicitly integrating $\delta(e_T)$ over our epitope density distribution $f(e_T)$.
Instead we need only calculate \eTmin, then $\left<e_T\right>$, and finally $\left<\delta\right>$.

\subsection{Calculations with immobile fraction}
Analysis of the model with non-zero immobile epitope fraction ($\eta > 0$) is considerably more difficult than the case of completely mobile epitopes, even with $\alpha = 0$.
To calculate $\delta$ we must now numerically solve a system of five, rather than three, algebraic equations.
Additionally, the simplifications of the previous section no longer apply, so we must find the bounding ligand and epitope concentrations for adhesion using numerical root-finding, and we must calculate $\left<\delta\right>$ by explicit numerical integration of $\delta(e_T)$ over the distribution $f(e_T)$.

When $\eta > 0$, direct solution of our system of five constraint equations will give non-physical results for those cases in which the immobile epitopes themselves are dense enough to drive adhesion.
Because the immobile epitope density is assumed constant over the cell surface, this case results in a divergent contact area $\delta$, as demonstrated in Fig.~\ref{fig:immobile_explain}.
In our application, this phenomenon only occurs at very high total epitope densities $e_T$, so it makes only a small contribution in our typical integrations over $f(e_T)$, but we must handle it carefully to avoid numerical difficulties.
The value of $\eta$ which leads to this divergence, $\eta_{\rm div}$ can be found by solving our conservation equation for immobile epitopes inside the contact region (Eq.~\ref{eqn:iin_constraint}) with $b_{01} = \beta$, yielding
\begin{equation}
\eta_{\rm div} = \beta \frac{1+2 K_I L (1 + K_{b1} r_\in)}{2 K_{b1} K_I L e_T r_\in}.
\end{equation}
When $\eta > \eta_{\rm div}$, we set $\delta = 0.5$, consistent with a cell completely flattened against the surface.
Similarly, we set $\delta = 0.5$ whenever direct solution of the equations would yield a larger value for $\delta$.
Altering this maximum value of $\delta$ has very small influence on our results (data not shown), as in our experiments the vast majority of cells have only a small fraction of their area adhered.

In our analyses, we consider fits with a constant immobile fraction, and fits in which the immobile fraction $\eta$ is a function of either $\delta$ or the fraction $\kappa$ of target cell epitopes that are cross-linked by ligand:
\begin{equation}
\kappa = \frac{2 \delta (e_{2\in} + b_{20} + h_{\in} + b_{11}) + 2 (1-\delta) (e_2 + h)}{e_T}.
\end{equation}
In both cases we consider a linear dependence of $\eta$ on $\delta$ or $\kappa$:
\begin{equation}\label{eq:eta_delta}
\eta = 0.13 + s_\delta \delta,
\end{equation}
or
\begin{equation}\label{eq:eta_kappa}
\eta = 0.13 + s_\kappa \kappa.
\end{equation}
Here 0.13 is the experimentally observed immobile epitope fraction in the absence of ligand~\cite{bib:Dustin2007}.
Eq.~\ref{eq:eta_delta} or~\ref{eq:eta_kappa} represents an additional constraint to our previous five, to account for the additional free variable $\eta$.
There may be multiple self-consistent solutions of our expanded system of six constraint equations (see Fig.~\ref{fig:immobile_explain}).
Physically, we expect the cell to adopt the solution corresponding to smaller $\eta$, as the cell begins in a state with minimal epitope immobility.
In our calculations we always adopt the smallest possible solution.

\section{Results}

Experiments were previously carried out to characterize alefacept-mediated bridging of CD2 epitopes on Jurkat T-cells to fluorescently-labeled Fc$\gamma$RIIIb
receptors on supported bilayers~\cite{bib:Dustin2007}.
The alefacept concentration $L$ was varied from 1 nM to 100 $\mu$M and the fraction of cells bound to the bilayer, the average size of the contact area, and the average number of bonds in the contact area were determined.
These types of measurements were made at two different Fc$\gamma$RIIIb densities $r_T$ on the bilayer: 1200 $\microm^{-2}$ and 625 $\microm^{-2}$.
Here we fit several models of increasing complexity to this data.
Using these models, we then consider the requirements for alefacept-mediated T-cell adhesion, deriving compact expressions for the limiting alefacept concentrations.
Finally, we consider the effect of background nonspecific IgG on adhesion.

\begin{figure}
\centering
\includegraphics{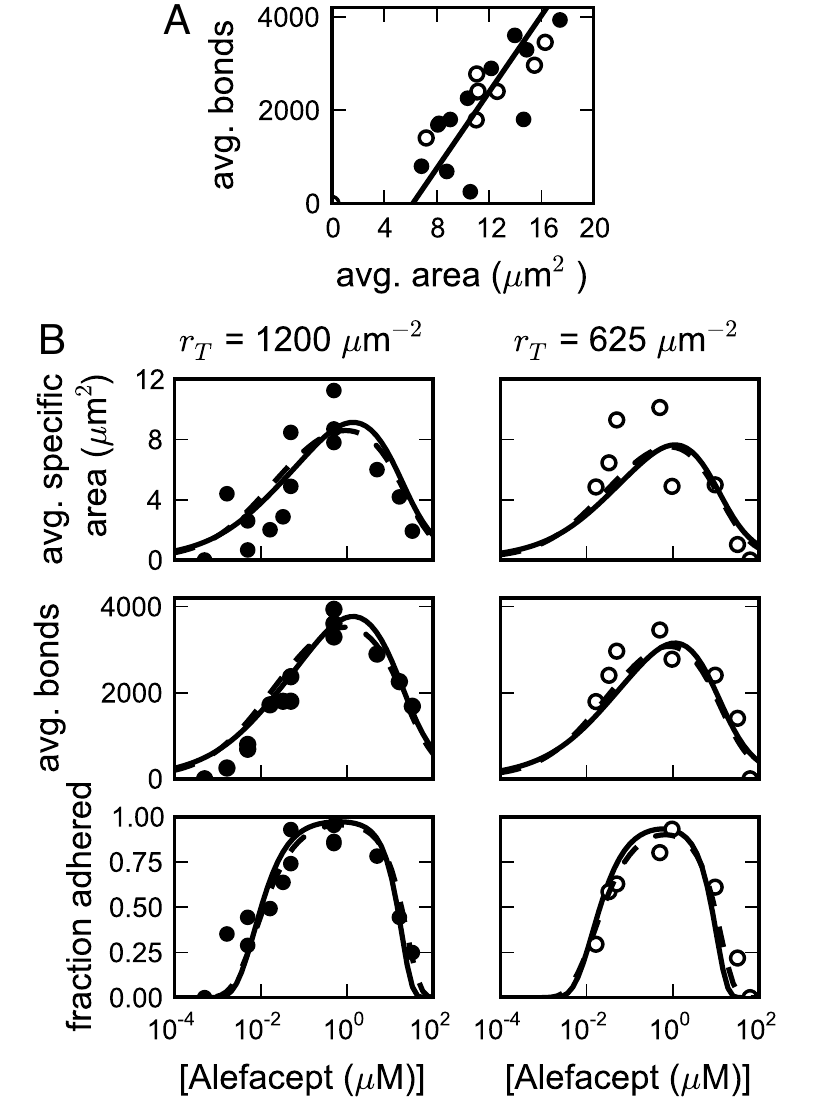}
%\subfigure[]{
%\includegraphics[]{bonds_v_area_homo}}
%\subfigure[]{\includegraphics[]{fit_data}}
\caption{A: Experimental data on average bridging bond count versus average contact area. Once the non-specific adhesion offset $A_{\rm off}$ is removed, a linear relationship is evident. 
We plot the model with $\beta = 410 / \microm^2$ (line).  
B: Fits to contact area, bridging bond, and fraction-adhered data for the best-fit model with all T-cell epitopes  mobile (solid lines) and the best-fit model with epitope immobility $\eta$ a linear function of the contact area $\delta$ (dashed lines). 
\label{fig:hetero_fits}}
\end{figure}

In Fig.~\ref{fig:hetero_fits}A the bridging bond data is plotted against the area of the contact region.
From this data we extract a best-fit value for the non-specific contact area of $\Aoff = 6.17\,\microm^2$, with a  95\% confidence interval of 2.8--8.5 $\microm^2$.
In all our subsequent fits, this value is subtracted from the contact area data, so we only fit the specific contact area.

\subsection{Fit of fully mobile model}

\begin{table*}
\caption{Fit results (best-fits and 95\% confidence intervals)\label{tbl:fit_results}}
\begin{tabular*}{\textwidth}{@{\extracolsep{\fill}}ccccccccccc}
model & $\chi^2$ & $\beta\,(\microm^{-2})$ & $K_{b1}\,(\microm^2)$ & $K_{b2}\,(\microm^2)$& $\eta$ & $s_\kappa$ & $s_\delta$  & $k$ & $\lambda$ $(\times 10^3)$ & $\left<E_T\right>$ $(\times 10^3)$\\
\hline
$\eta = 0$ & 1.658 & 413 & 1.3 & 54 &&&& 1.9 & 6.2 & 5.5 \\
                 &            & 364--487      &  0.3--3.0      & 23--108 &&&&     1.7--2.3     & 5.1--8.3       & 4.5--7.3\\
$\eta$ = constant  & 1.658 & 413 & 1.3 & 55 & 0 &&& 1.9 & 6.3 & 5.5\\
                &            & 368--495      & 0.4--2.6       & 26--102              & 0--0.01      & & & 1.7--2.2 & 5.2--9.0 & 4.6--8.0\\
$\eta = 0.13 + s_\kappa \kappa$ & 1.663 & 414 & 1.3 & 55 && 0 && 1.9 & 7.2 & 6.4\\
                                                    && 362--483 & 0.4--2.6 & 25--94 & & 0--0.06 && 1.7 -- 2.2 & 5.9--9.7 & 5.3--8.6\\
$\eta = 0.13 + s_\delta \delta$ & 1.432 & 411 & 0.7 & 5.5 & && 42 & 1.5 & 21 & 19\\
                                                  &         & 361--487 & 0.3--1.6 & 2.2--17.4 &&& 26--56 & 1.2--1.8 & 11--36& 10--32\\

\end{tabular*}
\end{table*}

We first considered an adhesion model with free diffusion of all Fc$\gamma$RIIIb receptors on the bilayer and all CD2 epitopes on the T cell.
Simultaneous nonlinear least-squares fits of this model to the data are shown by the solid lines in Fig.~\ref{fig:hetero_fits}B.
We weighted the experiments so that both the bond and contact area data went between zero and one, and we fit five parameters: the bridging bond density $\beta$, the equilibrium constants $K_{b1}$ and $K_{b2}$, and the Weibull distribution parameters $k$ and $\lambda$.
Table~\ref{tbl:fit_results} lists the best-fit values of the free parameters, along with 95\% confidence intervals.
The best-fit value for the average number of epitopes per cell was roughly 5,500, which is much smaller than the average value of 64,000 determined from flow cytometry (Fig.~\ref{fig:flowcyt}).
A reasonable fit could not be obtained when the distribution of epitopes per cell was taken directly from the flow cytometry data (data not shown).

One counter-intuitive property of the fits is that at high and low ligand concentrations it appears as if there is a slow decline in the average contact area and number of bridging bonds long after the number of bound cells has gone to zero.
In these cases, the average contact area is being calculated over the miniscule fraction of cells that are adhered.
For example, with $r_T = 1200 \microm^{-2}$, at $L = 10^{-4}\,\mu\text{M}$ an average of 125 bonds per adhered cell is predicted, but this involves only a fraction $10^{-27}$ of the total cells.
In the experiments, only a few hundred cells were sampled per data point, so the tails of the Weibull distribution are a poor description of the cell population.
Whether the Weibull distribution is a reasonable description of the epitope density on a target cell population \emph{in vivo} is an open question.
We have also considered a lognormal distribution.
Although it can fit the flow cytometry data in Fig.~\ref{fig:flowcyt} well, when it is used to analyze the data in Fig.~\ref{fig:hetero_fits},  it predicts that as the ligand concentration decreases, the average contact area and number of bridging bonds go through minima and then rise, yielding a very poor fit to this data (results not shown).
Thus, for the lognormal distribution (and possibly other distributions) the average number of bonds in the contact region can increase as the ligand concentration goes to zero. 

This model with freely diffusing CD2 epitopes dramatically underestimates the amount of CD2 present on the T cells.  Therefore we consider more complex models incorporating immobile CD2 epitopes in the following sections.

\subsection{Fit of models with epitope immobility}

In prior experiments on Jurkat T cells, it was observed that 13\% of CD2 epitopes were immobile in the absence of ligand~\cite{bib:Dustin2007}.
It has further been observed in T cells that cell stimulation may increase CD2 immobility~\cite{bib:Zhu2006}.
Thus we extended our mathematical model to include potential CD2 epitope immobility and fit several such models to the data.

We first considered a model in which the immobile epitope fraction $\eta$ was a fit constant.
In this case, the best-fit value of $\eta$ was found to be zero, yielding an identical fit to the fully mobile case.
This motivated us to consider models in which the immobile fraction was a function of the fraction $\kappa$ of epitopes cross-linked by ligand or the fractional area $\delta$ of the specific contact region.

When we fit a model in which the immobile fraction was a linear function of $\kappa$  (restricting our search to slopes $\geq$ 0), the best-fit value for the slope of that function was 0, yielding a constant immobile fraction of 0.13.
The resulting model fit was slightly worse than the completely mobile model.

The dashed curves in Fig.~\ref{fig:hetero_fits}B show the results from a fit with the immobile fraction a linear function of the specific contact area $\delta$.
In this case, the fit is somewhat improved, and the best-fit estimate of total epitope count per cell is driven upward to roughly $19,000$.
This estimate of $E_T$ is still only a about a third of the value inferred from the flow cytometry data, but it is much closer than the other models.
The best-fit function for $\eta$ is $\eta = 0.13 + 43 \delta$, so the immobile fraction increases very rapidly with cell adhesion.
Note that in our data, the largest average specific contact area seen is roughly $12\,\microm^2/\Acell = 0.015$ so that only small values of $\delta$ are typically explored, and adding higher-order terms to $\eta(\delta)$ yields negligible improvement in the fit (data not shown).

\subsection{Requirements for adhesion}

For drug design, an important consideration is what combinations of ligand concentration and target cell epitope count will yield binding.
The curves in Fig.~\ref{fig:nonspecific}A separate the region where more than 50\% of cells are adhered (inside each curve) from the region where less than 50\% are adhered for three different scenarios.
The outermost thick solid curve is the predicted separation curve for the parameters obtained from the fit with all receptors mobile (that shown by the solid lines in Fig.~\ref{fig:hetero_fits}B).
From this curve we can see that the minimal ligand concentration for adhesion $L_-$ is inversely proportional to the square of the epitope density, i.e.\ the bottom portion of the curve has a slope of approximately negative two.

From the complete set of equations for the model with $\eta = 0$ and $\alpha = 0$, we can obtain simple approximations for $L_-$ and $L_+$ for a fixed target cell epitope density $e_T$:
\begin{equation}\label{eqn:Lminus}
L_- \approx \frac{\beta}{e_T^2 K_E K_{b2} K_x r_T \sigma_E^2 \sigma_R}
\end{equation}
and
\begin{equation}\label{eqn:Lplus}
L_+ \approx \frac{e_T K_{b1} r_T \sigma_E \sigma_R}{\beta K_R}.
\end{equation}
To approximate $L_-$, we assume that, at the lowest ligand concentrations that mediate adhesion, all bridging bonds arise from epitopes bound bivalently (so $K_{b1} = 0$).
For the $L_+$ approximation, we assume that, at the highest ligand concentrations that mediate adhesion, all ligands bound to epitopes are bound singly (so $K_x = K_{b2} = 0$).
As seen in Fig.~\ref{fig:nonspecific}A, Eq.~\ref{eqn:Lminus} and~\ref{eqn:Lplus} closely predict the ligand concentration of 50\% adhesion, when we replace $e_T$ by the average epitope density.
Our approximate expression for $L_-$ suggests that ligands similar to alefacept can achieve considerable selectivity in epitope density on adhered cells, because $L_-$ falls with the square of the epitope density.
Further, since the ligand-epitope cross-linking constant ($K_x$) is proportional to the ligand-epitope binding constant ($K_E$), Eq.~\ref{eqn:Lminus} implies that $L_-$ falls inversely with the square of $K_E$, suggesting that a good strategy for lowering $L_-$ is to develop ligands with higher $K_E$ values.

The dashed curve in Fig.~\ref{fig:nonspecific}A bounds the region of adhesion for the best-fit model in which the immobile fraction $\eta$ is a linear function of the contact area $\delta$.
Again, the minimal ligand concentration for adhesion falls as the square of the epitope density.
Our expressions for $L_-$ and $L_+$ (Eq.~\ref{eqn:Lminus} and~\ref{eqn:Lplus}) are no longer good approximations in this case, but the dependence of $L_-$ on the ligand-epitope binding constants $K_E$ and $K_x$ is the same.
This is illustrated by the dotted curve, which bounds the region of adhesion for the same model and parameters, with the exception that $K_E$ and $K_x$ have been divided by 10.
As expected from Eq.~\ref{eqn:Lminus}, the minimal ligand concentration for adhesion has increased by a factor of 100.

\subsection{Adhesion inhibition by non-specific IgG}

\begin{figure}
\centering
\includegraphics{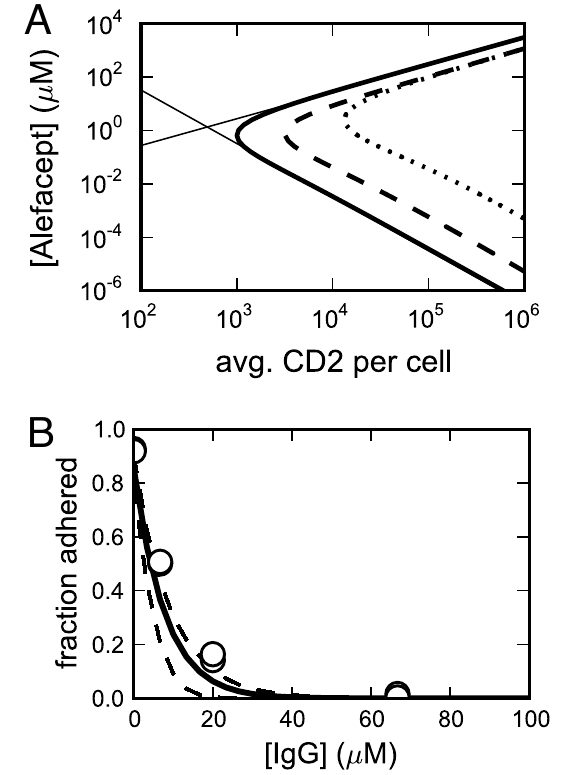}
%\subfigure[]{\includegraphics[]{LvE_hetero}}
%\subfigure[]{\includegraphics[]{nonspecific}}
\caption{A: Curves shown enclose the region of greater than 50\% cell adhesion, for the best-fit all-mobile model (thick solid), the best fit model with $\eta(\delta)$ (dashed), and the $\eta(\delta)$ model with ligand-epitope binding constants $K_E$ and $K_x$ each divided by 10 (dotted). The thin solid lines show our approximations for the bounding ligand concentrations $L_-$ and $L_+$. B: Experimental data on the inhibition of adhesion by non-specific IgG (open circles), compared with predictions from our model with the immobile epitope fraction a function of the contact area. The black line is from the best-fit model, while the dashed lines denote 95\% confidence intervals from our bootstrap parameter uncertainties. \label{fig:nonspecific}}
\end{figure}

In vivo, alefacept-mediated adhesion depends not only on the concentration of alefacept and the density of CD2 on the T cells, but also on the presence of background nonspecific IgG, which can bind the NK cell receptor and inhibit adhesion.
To test the sensitivity of adhesion to background concentrations of nonspecific IgG, the percentage of cells bound to the bilayer was determined in the presence of 500 nM of the ligand (alefacept) and differing concentrations $G$ of purified human IgG, with a receptor concentration in the bilayer of 450 $\microm^{-2}$~\cite{bib:Dustin2007}.
For inhibition to be significant, we expect that $K_G G$ must be greater than 1, where in these experiments the IgG binding constant $K_G$ is equal to our $K_R$.
In Fig.~\ref{fig:nonspecific}B, an inhibition of 50\%  is achieved with an IgG concentration of about 7 $\mu$M, suggesting that $K_R \geq 1.5\times 10^5$ M$^{-1}$.
(In our fits we took $K_R=1.0\times 10^{6}$ M$^{-1}$~\cite{bib:Galon1997,bib:Chesla2000,bib:Maenaka2001}.)
Fig.~\ref{fig:nonspecific}B shows good agreement between our predicted inhibition curve using the best-fit parameters for the $\eta(\delta)$ model and the experimental data, providing further validation of our model.

\section{Discussion}

We have developed an equilibrium model for the ligand-mediated adhesion of cells to surfaces.
Our model incorporates potential heterogeneities in target cell epitope density, immobility of epitopes, and the possibility that ligand binding or adhesion alters the immobile epitope fraction.
We have applied our model to experiments on the alefacept-mediated adhesion of Jurkat T cells expressing CD2 to bilayer membranes containing the receptor Fc$\gamma$RIIIb, a close relative of the relevant receptor on natural killer cells.
We find that our data are best described by a model in which the immobile epitope fraction is a function of the contact area between the target cell and bilayer.
Nevertheless, our best-fit model still underestimates the epitope density on Jurkat T cells, perhaps indicating that other factors influence CD2-mediated adhesion and opening a direction for future study.

Our results also suggest general guidelines for the design of immuno-adhesive molecules.
We find that, for bivalent ligands, the minimal ligand concentration $L_-$ required for adhesion is inversely related to the square of the target cell epitope density, illustrating the potential selectivity of these ligands.
We also show that $L_-$ is a quadratic function of the epitope-ligand binding constants, even for our more complex models, suggesting that tuning this interaction may be a fruitful route for drug design.

It is instructive to compare our fit parameters with with those from previous investigations of Jurkat T cell adhesion.
A previous analysis of adhesion to bilayers containing the natural CD2 binding partner, CD58, found a non-specific contact area of 7.6 $\microm^2$, similar to the 6.17 $\microm^2$ we find~\cite{bib:Shao2005}.
That analysis found the density of bridging bonds $\beta$ at equilibrium to be approximately 1000 $\microm^{-2}$, substantially greater than our value of 400 $\microm^{-2}$.
This is unsurprising, because the direct CD2-CD58 interaction draws the cell closer to the bilayer, leading to a larger repulsive force which requires more bonds to overcome.
Our value for the 2D association constant $K_{b1}$ between the alefacept-epitope complex and CD16b are of order 1 $\microm^2$.
This value is similar to previous results for the 2D association constant between CD2 and CD58, which range from 0.1 to 0.9 $\microm^2$~\cite{bib:Dustin1997, bib:Zhu2006, bib:Zhu2007}.  Thus our fit parameters are consistent with those from the literature.

Our data are best described by a model in which the level of immobilized CD2 on the T cell is proportional to the contact area, rather than the degree of CD2 cross-linking.
This suggests that signaling from isolated cross-linked CD2 pairs may be less effective than signaling from larger aggregates or from regions of high CD2 density, such as the contact region.
A similar effect is seen in signaling from the immune receptor Fc$\epsilon$RI on rat basophilic leukemia cells, in which receptor dimers signal weakly compared to larger aggregates~\cite{bib:Fewtrell1980}.
Furthermore, recent experiments have observed signaling in Jurkat T cells adhering to bilayers presenting CD58, the natural binding partner of CD2~\cite{bib:Kaizuka2009}.

In our system, comparing the bridging bond density $\beta$ of approximately 400 $\microm^{-2}$ with the initial cellular CD2 density of roughly 80 $\microm^{-2}$, we see that the density increase in the contact region is roughly a factor of 5.
This increases the fraction of CD2 nearest neighbors within a given distance by the same factor of 5 (Eq.~\ref{eqn:distances}), substantially increasing the probability of interaction.
Thus the increased density of CD2 in the contact region may have a large effect in promoting interactions between CD2 molecules and driving signaling.
When alefacept adheres T cells to NK cells in vivo, Fc$\gamma$RIIIa receptors on the NK cell will be similarly concentrated in the contact region, and this concentration may contribute to the signal that drives NK-mediated killing of target cells.
Moreover, we expect the repulsive force to be greater between cells than between a cell and a bilayer, so the in vivo bridging bond density is probably larger, leading to greater in vivo receptor concentration than seen in our experiments.

In summary, we have developed an equilibrium model for the immunoadhesin-mediated adhesion of cells to surfaces.
Our analysis suggests guidelines for the design of therapeutic immunoadhesins.
Furthermore, applying our model to experiments on Jurkat T cells suggests that an active cellular process may be increasing CD2 immobility in response to alefacept-mediated surface adhesion.

\begin{acknowledgments}
This work was supported by the Department of Energy through contract W-7405-ENG-36 and the National Institutes of Health through grants R37-GM035556 and R56-AI44931. Further support was provided by the National Science and Engineering Research Council and the Mathematics of Information Technology and Complex Systems National Centre of Excellence.
\end{acknowledgments}

\bibliography{refs}

\clearpage

\appendix

\renewcommand{\theequation}{S.\arabic{equation}}
\renewcommand{\thesubsection}{S.\arabic{subsection}}
\renewcommand{\thefigure}{S.\arabic{figure}}
\setcounter{figure}{0}
\setcounter{equation}{0}

\onecolumngrid

\section*{Supporting Material:\\
\ourtitle}

\subsection{Full model}

The full set of equilibrium relations among the molecular complexes considered in our model is given by Eq.~\ref{eqn:equil_first} through~\ref{eqn:equil_last}, which express all complex concentrations as functions of $e$, $r$, $i$, and $i_{\rm in}$:
\begin{gather} \label{eqn:equil_first}
e_{\rm in} = \sigma_E e,\\
r_{\rm in} = \sigma_R r,\\
e_1 = 2 K_E   L   e,\\
e_2 = K_x e_1   e / 2,\\
e_{1\rm in} = 2   K_E   L   e_{\rm in},\\
e_{2\rm in} = K_x   e_{1\rm in}   e_{\rm in} / 2,\\
b_{10} = K_{b10}   e_{1\rm in}   r_{\rm in},\\
b_{20} = K_{b20}   e_{2\rm in}   r_{\rm in},\\
r1 = K_R   L   r,\\
r_{1\rm in} = K_R   L   r_{\rm in},\\
i_1 = 2 K_I   L   i,\\
i_{1\rm in} = 2 K_I   L   i_{\rm in},\\
h = K_{HE}   e_1   i,\\
h_{\rm in} = K_{HE}   e_{1\rm in}   i_{\rm in},\\
b_{01} = K_{b01}   i_{1\rm in}   r_{\rm in}, \text{ and}\\
b_{11} = K_{FH}   h_{\rm in}   r_{\rm in}. \label{eqn:equil_last}
\end{gather}
Note that the above equations do not involve any of the underscored equilibrium constants indicated in Fig.~\ref{fig:rxns} of the main text.
This is because enforcing detailed balance around all loops in the reaction diagram introduces six constraints on the equilibrium constants, which we use to eliminate the underlined constants:
\begin{gather}
K_x K_{b20} = \underline{K_x} K_{b10},\\
K_E K_{b10} = \underline{K_E} K_R,\\
K_{b10} \underline{K_{FE}} = K_{FH} K_{HE},\\
K_I K_{b01} = \underline{K_I} K_R,\\
K_E K_{HE} = K_I \underline{K_{HI}},\text{ and}\\
K_I K_{b01} \underline{K_{FI}} = K_{FH} K_{HE} K_E.
\end{gather}

\subsection{Requirements for adhesion}

As described in the main text, for the model with all receptors mobile, we can find equations for the bounds on ligand concentration $L$ and epitope density $e_T$ for adhesion by setting $\delta = 0$ in Eq.~\ref{eqn:epitope_constraint}, \ref{eqn:receptor_constraint}, and \ref{eqn:beta_constraint}.

In the model with all receptors mobile, to find the minimal epitope density \eTmin for adhesion, we solve the quadratic equation
\begin{equation}\label{eqn:eTmin}
a_e\,e_T^2 + b_e\,e_T + c_e = 0
\end{equation}
where the coefficients are given by:
\begin{gather}
a_e = {K_{b2}^\prime}^2 K_E K_x L r_T^2,\\
b_e = 2 K_E L r_T \left[-4 {K_{b1}^\prime}^2 K_E L\, r_T + K_{b1}^\prime K_{b2}^\prime  r_T (1 + 2 K_E L) - 2 K_{b2}^\prime K_x \beta (1 + K_R L + K_G G)\right],\text{ and}\\
c_e = \beta (1 + K_R L + K_G G) \left\{4 K_E L \left[K_{b1}^\prime (r_T + 2 K_E L r_T)  + \beta K_x (1 + K_R L + K_G G)\right] - K_{b2}^\prime r_T (1 + 2 K_E L)^2 \right\},
\end{gather}
with $K_{b1}^\prime \equiv \sigma_R \sigma_E K_{b1}$ and $K_{b2}^\prime \equiv \sigma_R \sigma_E^2 K_{b2}$.
Eq.~\ref{eqn:eTmin} has two solutions, the larger of which is \eTmin.

To find the limits on ligand concentration for adhesion, we solve the cubic equation
\begin{equation}\label{eqn:Llim}
a_L\,L^3 + b_L\,L^2 + c_L\,L + d_L = 0
\end{equation}
where the coefficients are given by:
\begin{gather}
a_L = 4 \beta K_E K_R   (2 r_T K_{b1}^\prime K_E - r_T K_{b2}^\prime K_E + \beta K_R K_x),\\
\begin{align}
b_L = -4 K_E \big\{&\beta r_T \left[K_{b2}^\prime (K_E + K_R) - K_{b1}^\prime r_T (2 K_E + K_R) - 2 K_R K_x \beta\right]\\ &+
                e_T r_T (2 {K_{b1}^\prime}^2 K_E r_T - K_{b1}^\prime K_{b2}^\prime K_E r_T + K_{b2}^\prime K_R K_x \beta)\big\},\nonumber
                \end{align}\\
c_L = 2 r_T K_{b1}^\prime K_E   (2 \beta + e_T r_T K_{b2}^\prime) + 4   \beta^2   K_E K_x
            + e_T^2   r_T^2   {K_{b2}^\prime}^2   K_E   K_x
            - r_T \beta K_{b2}^\prime   \left[K_R + 4   K_E   (1 + K_x e_T)\right],\text{ and}\\
d_L = -r_T \beta K_{b2}^\prime.
\end{gather}
Eq.~\ref{eqn:Llim} has either two positive solutions, which are $L_-$ and $L_+$, or no positive solutions, in which case adhesion is not possible for any ligand concentration.

\subsection{Sensitivity to $K_x$}

\begin{figure}[ht]
\centering
\includegraphics[width=3in]{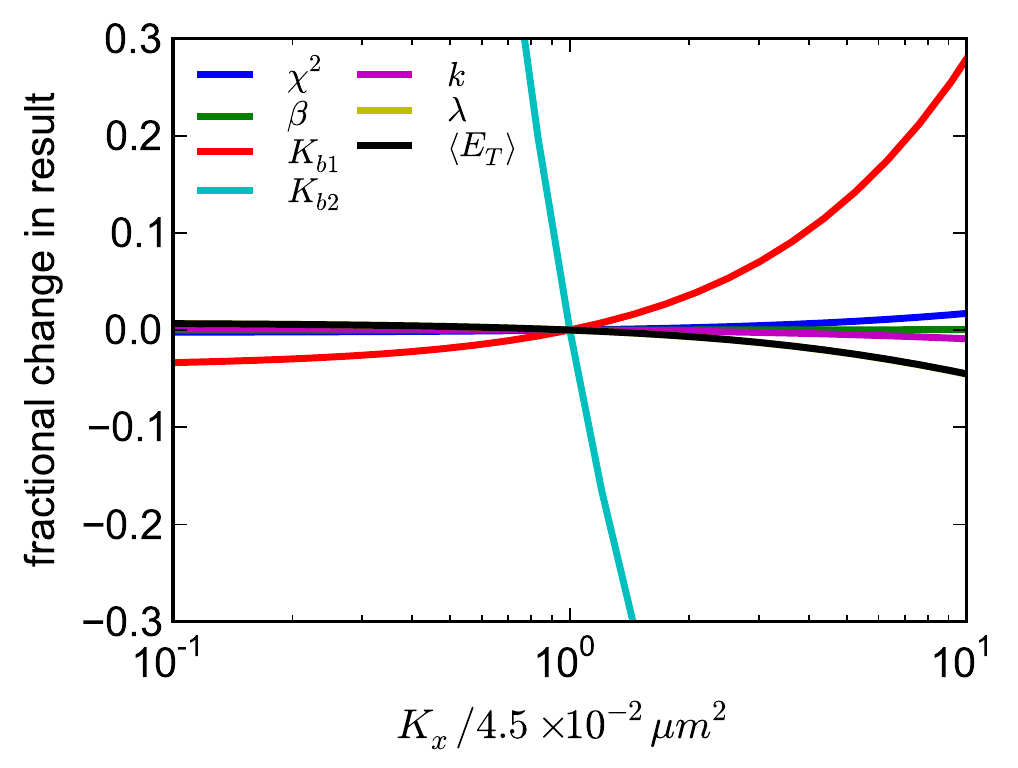}
\caption{Sensitivity to $K_x$. The change in best-fit parameter values for the completely mobile model is plotted as $K_x$ varies from our inferred value of $4.5 \times 10^{-2}\, \microm^2$. With the exception of $K_{b2}$, changing $K_x$ by a factor of 10 up or down from this value changes the best-fit results by less than 30\%. The best-fit value of $K_{b2}$, on the other hand, is inversely proportional to the value of $K_x$ assumed.\label{fig:changing_Kx}}
\end{figure}

\clearpage

\subsection{Model solutions and divergence}

\begin{figure}[ht]
\centering
\includegraphics[]{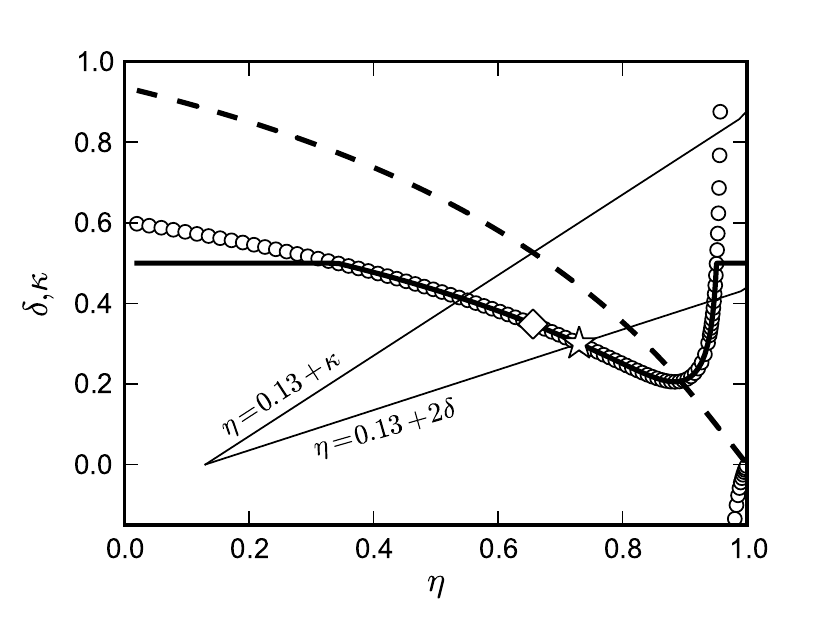}
\caption{Illustrative solutions of our model as a function of immobile fraction $\eta$. We show the solution for $\delta$ as a function of $\eta$ for a particular set of model parameters (open circles).
Note divergence at $\eta_{\rm div} \approx 0.95$. 
We show the filtered solution for $\delta$, accounting for the divergence and fixing $\delta\le0.5$ (thick solid line), and the fraction $\kappa$ of cross-linked epitopes as a function of $\eta$ (dashed line).
We also show two potential models with $\eta$ constrained to be a linear function of $\kappa$ or $\delta$ (thin solid lines). The solution for $\delta(\eta)$ with $\eta = 0.13 + 2 \delta$ is shown by the star, while the solution for $\delta(\eta)$ with $\eta = 0.13 + \kappa$ is shown by the large diamond. \label{fig:immobile_explain}}
\end{figure}

\end{document}